\documentclass[seceq]{ptptex}

\usepackage{graphicx}

\notypesetlogo                       

\providecommand{\LyX}{L\kern-.1667em\lower.25em\hbox{Y}\kern-.125emX\@}

\title{
Does $ \Sigma -\Sigma -\alpha  $ Form a Quasi-Bound
State?
}

\author{
Htun Htun \textsc{Oo}$ ^{1} $, Khin Swe  \textsc{Myint}$ ^{1} $,
Hiroyuki \textsc{Kamada}$ ^{2} $ and  Walter  \textsc{Gl\"ockle}$ ^{3} $
}

\inst{
 $ ^{1} $ Department of Physics, Mandalay University,
Mandalay, Myanmar \par
 $ ^{2} $ Department of Physics, Faculty of Engineering,
Kyushyu Institute of Technology, Kitakyushyu 804-8550, Japan \par
 $ ^{3} $ Institute for theoretical Physics II, Ruhr-University
Bochum, D-44780 Bochum, Germany
}

\abst{
We have investigated the possible existence of a quasi-bound state
for the $\Sigma -\Sigma -\alpha $ system in the framework of Faddeev
calculations. We are particularly interested in the state of total
iso-spin T=2, since for an inert $\alpha$ particle there is
no strong conversion to $\Xi -N-\alpha  $ or 
$\Lambda -\Lambda -\alpha $ possible. A $\Sigma -\alpha $
optical potential based on Nijmegen
model D and original $\Sigma -\Sigma $ interactions of the series
of Nijmegen potentials NSC97 as well a simulated Gaussian type versions
thereof are used. Our investigation of the 
$\Sigma -\Sigma -\alpha $ system leads to a quasi bound state where, 
depending on the potential
parameters,  the energy ranges between -1.4 and -2.4 MeV and the level
width is about 0.2MeV .
}

\begin{document}

\maketitle

\section{Introduction}

Strangeness S=-2 hypernuclei provide information on baryon-baryon
forces in the state of S=-2. Only three nuclei have been identified
so far, $ ^{10}_{\Lambda \Lambda }Be $
 \cite{ref1}, $^{6}_{\Lambda \Lambda }He $
\cite{ref2} 
and 
$^{13}_{\Lambda \Lambda }B $
\cite{ref3}. The challenge
is to understand their binding energies and decay properties. These
nuclei are especially interesting since the S=-2 two-baryon system
is rich in structure due to the conversions between $ \Lambda \Lambda  $
, 
$\Xi N $ 
and $\Sigma \Sigma $. Baryon-baryon forces for
S=0,-1 and -2 are being investigated in the meson exchange
picture\cite{ref4,ref5,ref6,ref7} 
or using quark models\cite{ref8}. While there is a wealth of data for
S=0, which allows to fix force parameters, the situation is still
much open in the S=-1 and -2 sectors. Therefore additional information
is needed. In this study we would like to focus on the system
$\Sigma -\Sigma -\alpha $ in the state of total iso-spin T=2. If the 
$\alpha$-particle
would be inert, that system could not convert to $\Xi -N-\alpha $
or $\Lambda -\Lambda -\alpha $. 
Therefore in case the forces
would be strong enough, there might exist a low lying state with a
small width. The width would be caused by $ \Lambda -\Sigma $ 
conversion leading , for instance, to $\Sigma +\Lambda
+N+(^{3}He/^{3}H) $, where $(N+^{3}He/^{3}H) $
 is in a state of total iso-spin T=1
or into  $ \Sigma +\Lambda +^{4}He^{*} $ (T=1).

We investigate that system $ \Sigma -\Sigma -\alpha $ under effective
simplifying assumptions. The $ \Sigma -\alpha  $ interaction is
modeled via an optical potential based on the Nijmegen model D and
the $ \Sigma -\Sigma  $ interaction in the state of total iso-spin
T=2 is taken either directly as a meson theoretical Nijmegen potential
of the type NSC97\cite{ref6} or a simulated version thereof of the Gaussian
type\cite{ref9}. That 3-body system is solved precisely in the Faddeev
scheme.

In section \ref{sec2} we derive the Faddeev equations for the 
$ \Sigma -\Sigma -\alpha  $
system. The numerical results are displayed in section \ref{sec3}. 
The Appendix
shows technical details of the formulation. We end with a brief
summary  of our work  
and  an  outlook in section \ref{sec4}.

\section{ The Faddeev Equations for the System \( \Sigma -\Sigma -\alpha  \)}
\label{sec2}

We assume the existence of a quasi-bound state where the width of
that state is caused by the (absorptive) imaginary part of the effective
\( \Sigma -\alpha  \) potential. If the \( \alpha  \) particle would
be inert and the two \( \Sigma  \)'s couple to iso-spin T=2, the
system cannot convert into \( \Xi -N-\alpha  \) or \( \Lambda -\Lambda -\alpha  \).
We focus on such a state with T=2. The Schr\"odinger equation reads

\begin{equation}
\left( H_{0}+V_{\Sigma \Sigma }+V_{\Sigma _{1}\alpha }+V_{\Sigma _{2}\alpha }\right) \Psi =E\Psi 
\end{equation}

It is convenient to number the two \( \Sigma  \)'s as 1 and 2 and
the \( \alpha  \) particle as 3. Then the Schr\"odinger equation converted
into an integral equation reads

\begin{equation}
\Psi =\frac{1}{E-H_{0}}\left( V_{12}+V_{13}+V_{23}\right) \Psi 
\end{equation}

which suggests the decomposition

\begin{equation}
\Psi =\psi _{12}+\psi _{13}+\psi _{23}
\end{equation}

with 

\begin{equation}
\psi _{ij}=\frac{1}{E-H_{0}}V_{ij}\Psi 
\end{equation}

Because of the identity of the \( \Sigma  \)'s and the antisymmetry
of \( \Psi  \) the two Faddeev components \( \psi _{23} \) and \( \psi _{13} \)
are not independent and are related to each other by

\begin{equation}
\psi _{13}=-P_{12}\psi _{23}
\end{equation}

This then leads to two coupled equations 

\begin{equation}
\psi _{12}=\frac{1}{E-H_{0}}V_{12}\left( \psi _{12}+\left( 1-P_{12}\right) \psi _{23}\right) 
\end{equation}

\begin{equation}
\psi _{23}=\frac{1}{E-H_{0}}V_{23}\left( -P_{12}\psi _{23}+\psi _{12}\right) 
\end{equation}

In a standard manner one introduces the two-body t-matrices

\begin{equation}
t_{12}=V_{12}+V_{12}\frac{1}{E-H_{0}}t_{12}
\end{equation}

\begin{equation}
t_{23}=V_{23}+V_{23}\frac{1}{E-H_{0}}t_{23}
\end{equation}

and obtains the final set of two-coupled Faddeev equations

\begin{equation}
\psi _{12}=\frac{1}{E-H_{0}}t_{12}\left( 1-P_{12}\right) \psi _{23}
\label{eq.10}
\end{equation}

\begin{equation}
\psi _{23}=\frac{1}{E-H_{0}}t_{23}\left( -P_{12}\psi _{23}+\psi _{12}\right) 
\label{eq.11}
\end{equation}

The total state is then given as 

\begin{equation}
\Psi =\left( 1-P_{12}\right) \psi _{23}+\psi _{12}
\end{equation}

We solve that system in momentum space and use a partial wave decomposition.
To that aim we introduce two types of Jacobi momenta in terms of the
individual momenta \( \mathbf{k}_{i} \) , i=1,2,3;

\begin{equation}
\mathbf{p}_{3}=\frac{1}{2}\left( \mathbf{k}_{1}-\mathbf{k}_{2}\right) 
\end{equation}

\begin{equation}
\mathbf{q}_{3}=\frac{2m_{\Sigma }\mathbf{k}_{3}-m_{\alpha }\left( \mathbf{k}_{1}+\mathbf{k}_{2}\right) }{2m_{\Sigma }+m_{\alpha }}
\end{equation}

and 

\begin{equation}
\mathbf{p}_{1}=\frac{m_{\alpha }\mathbf{k}_{2}-m_{\Sigma }\mathbf{k}_{1}}{m_{\alpha }+m_{\Sigma }}
\end{equation}

\begin{equation}
\mathbf{q}_{1}=\frac{\left( m_{\alpha }+m_{\Sigma }\right) \mathbf{k}_{1}-m_{\Sigma }\left( \mathbf{k}_{2}+\mathbf{k}_{3}\right) }{m_{\alpha }+2m_{\Sigma }}
\end{equation}

The momenta \( \mathbf{p}_{3} \) ,\( \mathbf{q}_{3} \)
and \( \mathbf{p}_{1} \) ,\( \mathbf{q}_{1} \) are
adequate for the Faddeev amplitudes \( \psi _{12} \) and \( \psi _{23} \)
which are driven by the two-body t-matrices \( t_{12} \) and \( t_{23} \),
respectively. Related to these momenta are partial wave projected
basis states

\begin{equation}
\left| p_{3}q_{3}\left( l_{3}s_{3}\right) j_{3}\lambda _{3}\left( j_{3}\lambda _{3}\right) JM(11)2\right\rangle \equiv \left| p_{3}q_{3}\alpha _{3}\right\rangle 
\end{equation}

and 

\begin{equation}
\left| p_{1}q_{1}\left( l_{1}\frac{1}{2}\right) j_{1}\left( \lambda _{1}\frac{1}{2}\right) I_{1}\left( j_{1}I_{1}\right) JM(11)2\right\rangle \equiv \left| p_{1}q_{1}\alpha _{1}\right\rangle 
\end{equation}

The sequence of discrete quantum numbers denote orbital and spin angular
momenta, their intermediate couplings and the couplings to the total
3-body angular momentum J with magnetic quantum number M. Finally
(11)2 denotes the iso-spin coupling. Because of the identity of particles
1 and 2, \( (l_{3}+s_{3}) \) has to be even. This imposes the only
restriction to these intermediate quantum numbers.

By a standard procedure\cite{ref10} the coupled set of equations (\ref{eq.10}-\ref{eq.11})
is projected onto those basis states. It results

\begin{eqnarray}
\psi ^{\alpha _{3}}_{12}\left( p_{3}q_{3}\right)  & \equiv \left\langle p_{3}q_{3}\alpha _{3}\right. \left| \psi _{12}\right\rangle  & =\frac{2}{E-\frac{p^{2}_{3}}{2\mu _{3}}-\frac{q^{2}_{3}}{2M_{3}}}\sum _{\alpha _{1}}\int ^{\infty }_{0}dq_{1}q^{2}_{1}\nonumber \\
 &  & \int ^{1}_{-1}dxt_{12}\left( p_{3},\pi _{2},E-\frac{q^{2}_{3}}{2M_{3}}\right) \nonumber \\
 &  & G_{\alpha _{3}\alpha _{1}}\left( q_{3}q_{1}x\right) \psi ^{\alpha _{2}}_{23}\left( \pi '_{2,}q_{1}\right) 
\label{eq.19}
\end{eqnarray}

\begin{eqnarray}
\psi ^{\alpha _{1}}_{23}\left( p_{1}q_{1}\right)  & = & \frac{1}{E-\frac{p^{2}_{1}}{2\mu _{1}}-\frac{q^{2}_{1}}{2M_{1}}}\sum _{\alpha _{3}}\int ^{\infty }_{0}dq_{3}q^{2}_{3}\int ^{1}_{-1}dxt_{23}\left( p_{1},\pi _{3},E-\frac{q^{2}_{1}}{2M_{1}}\right) \nonumber \\
 &  & G_{\alpha _{3}\alpha _{1}}\left( q_{3}q_{1}x\right) \psi ^{\alpha _{3}}_{12}\left( \pi '_{3}q_{3}\right) \nonumber \\
 &  & -\frac{1}{E-\frac{p^{2}_{1}}{2\mu _{1}}-\frac{q^{2}_{1}}{2M_{1}}}\sum _{\alpha '_{1}}\int ^{\infty }_{0}dq'_{1}q'^{2}_{1}\int ^{1}_{-1}dxt_{23}\left( p_{1},\pi _{4},E-\frac{q^{2}_{1}}{2M_{1}}\right) \nonumber \\
 &  & G_{\alpha _{1}\alpha '_{1}}\left( q_{1}q'_{1}x\right) \psi ^{\alpha '_{1}}_{23}\left( \pi '_{4}q'_{1}\right) 
\label{eq.20}
\end{eqnarray}

The purely geometrical quantities \( G_{\alpha \alpha '} \) resulting
from recouplings are given in the Appendix. Further the shifted arguments
in the two-body t-matrices and the Faddeev amplitudes under the integrals
are given as

\( \pi _{2}=\sqrt{q^{2}_{1}+\rho ^{2}_{1}q^{2}_{3}+2\rho _{1}q_{1}q_{3}x} \)~~~~~~~~\( \pi _{3}=\sqrt{\rho ^{2}_{2}q^{2}_{1}+q^{2}_{3}+2\rho _{2}q_{1}q_{3}x} \) 

\( \pi '_{2}=\sqrt{\rho ^{2}_{2}q^{2}_{1}+q^{2}_{3}+2\rho _{2}q_{1}q_{3}x} \)~~~~~~~~
\( \pi '_{3}=\sqrt{q^{2}_{1}+\rho ^{2}_{1}q^{2}_{3}+2\rho _{1}q_{1}q_{3}x} \) 

\( \pi _{4}=\sqrt{\rho ^{2}q^{2}_{1}+q'^{2}_{1}+2\rho q_{1}q'_{1}x} \)~~~~~~~~~\( \pi '_{4}=\sqrt{q^{2}_{1}+\rho ^{2}q'^{2}_{1}+2\rho q_{1}q'_{1}x} \)

where \( \rho _{1}=\frac{1}{2} \),\( \rho _{2}=\frac{M_{\alpha }}{M_{\alpha }+M_{\Sigma }} \)
and \( \rho =\frac{M_{\Sigma }}{M_{\Sigma }+M_{\alpha }} \)

The reduced masses are 

\( \mu _{3}=\frac{1}{2}M_{\Sigma } \)

\( M_{3}=\frac{2M_{\alpha }M_{\Sigma }}{2M_{\Sigma }+M_{\alpha }} \)

\( \mu _{1}=\frac{M_{\alpha }M_{\Sigma }}{M_{\alpha }+M_{\Sigma }} \)

\( M_{1}=\frac{M_{\Sigma }\left( M_{\Sigma }+M_{\alpha }\right) }{2M_{\Sigma }+M_{\alpha }} \)

This is an infinite set of homogenous coupled integral equations in
two variables, which can be truncated since the two-body t-matrices
drop quickly in magnitude with increasing angular momenta.

\section{Results}
\label{sec3}

The set of coupled equations (\ref{eq.19}-\ref{eq.20}) 
are discretized in a standard
manner. We choose Gaussian quadrature points in the variables q and
x and spline interpolation for the variables \( \pi  \) under the
integrals. The two-body t-matrices are generated from the Lippmann
Schwinger equation again using Gaussian quadrature discretization.
We refer for numerical details to \cite{ref10,ref11}. The energy eigenvalue
E is determined as follows. The homogenous set of coupled equations
is schematically written as

\[
\eta (E)\psi =K(E)\psi \]

with \( \eta (E)=1 \) at the energy eigenvalue. Without knowing E,
one first determines \( \eta  \) and varies E such that finally 
\( \eta (E)=1 \).
The eigenvalue \( \eta  \) is determined either by a simple power
method or by a Lanczos type algorithm. For details see \cite{ref11,ref12}.

In order to demonstrate our numerical accuracy we would firstly like
to display results on the \( \Lambda \Lambda \alpha  \) system, where
for model forces we recalculated \( ^{6}_{\Lambda \Lambda }He \)
binding energies. The model forces are of Gaussian types and we refer
to \cite{ref13,ref14}. That system has been investigated before by Filikhin
et al. \cite{ref13} using Faddeev equations in configuration space and
by Myint et al. \cite{ref14} using a variational method based on a Gaussian
expansion. Our method is mathematically different from theirs. We
solved the Faddeev equations in momentum space which allows us to
treat any type of two body forces. We show in Table \ref{table1} 
\( ^{6}_{\Lambda \Lambda }He \)
binding energies for an increasing number of relative orbital angular
momenta within the \( \Lambda \alpha  \) and \( \Lambda \Lambda  \)
sub-systems. Our results are in very good agreement with the other
two methods which underlines the reliability of our treatment.

\begin{table}[tp]
\begin{center}
\begin{tabular}{|c|c|c|c|c|}
\hline 
\( l_{\Lambda \alpha } \)&
\( l_{\Lambda \Lambda } \)&
Ours&
Filikhin&
Myint\\
\hline
\hline 
0&
0&
-6.880&
-6.880&
-6.880\\
\hline 
0,1&
0&
-6.986&
-6.987&
-6.983\\
\hline 
0,1,2&
0,2&
-7.050&
-7.045&
-7.041\\
\hline 
0,1,2,3&
0,2&
-7.084&
-7.078&
-7.073\\
\hline
\end{tabular}
\caption{
Comparison of $ ^{6}_{\Lambda \Lambda }He $ binding energies
in MeV based on the model forces given in \cite{ref13} and \cite{ref14} for
an increasing number of partial waves.}
  \label{table1}
  \end{center}
\end{table}

Now we turn to the central topic, the \( \Sigma \Sigma \alpha  \)
system in the state of total iso-spin T=2. For the  \( \Sigma -\Sigma  \)
potential we use either the original Nijmegen potentials NSC97a,c,e
\cite{ref4} or the simulated Gaussian forms thereof \cite{ref9}. 
The latters
are given as

\begin{equation}
V(r)=\sum ^{2}_{i=1}V_{i}e^{-(r/\mu _{i})^{2}}
\label{eq.21}
\end{equation}

with the parameters shown in Table \ref{table2}. The \( \Sigma -\alpha  \) potential
is chosen to be complex to provide for absorptive processes, like
the ones mentioned in the introduction. We use the form

\begin{equation}
V_{\Sigma \alpha }(r)=\sum ^{2}_{i=1}V_{i}e^{-(r/\mu _{i})^{2}}+\sum ^{2}_{i=1}U_{i}e^{-(r/\mu _{i})^{2}}
\end{equation}

with the parameters \( V_{1}=-21.3 \) MeV, \( V_{2}=4.8 \) MeV for
the real part and \( U_{1}=4.07 \) MeV, \( U_{2}=-11.73 \) MeV for
the imaginary part. Further, one has \( \mu _{1}=1.3 \) fm and \( \mu _{2}=1.7 \)
fm.

The \( \Sigma -\alpha  \) potential has been constructed in the following
manner. An effective \( \Sigma -N \) potential was firstly derived
from the original Nijmegen model D interaction in the Brueckner framework.
Then that effective potential was expanded into a five-range Gausian
form. That potential was used in a generalized Hartree-Fock method to generate
the effective \( \Sigma -\alpha  \) potential\cite{ref15}. The imaginary
part arises due to \( \Lambda N \) to \( \Sigma N \) conversion.

\begin{table}[tp]
\begin{center}
\begin{tabular}{|c|c|c|}
\hline 
&
\( V_{1} \) (MeV)&
\( V_{2} \) (MeV)\\
\hline
\hline 
NSC97a ``sim''&
5274.2576&
-292.7193\\
\hline 
NSC97c ``sim''&
4886.7830&
-289.8740\\
\hline 
NSC97e ``sim''&
4588.5040&
-294.2548\\
\hline
\end{tabular}
\caption{
The potential strength parameters of the Gaussian form {}``sim''
given in Eq.(\ref{eq.21}) for the \( \Sigma -\Sigma  \) system. The range
parameters are \( \mu _{1}=0.37 \) fm and \( \mu _{2}=1.0 \) fm.
}
  \label{table2}
  \end{center}
\end{table}

\begin{figure}
\begin{center}
\vspace{0.3cm}
{\centering \resizebox*{1\textwidth}{0.6\textheight}{\rotatebox{-90}{\includegraphics{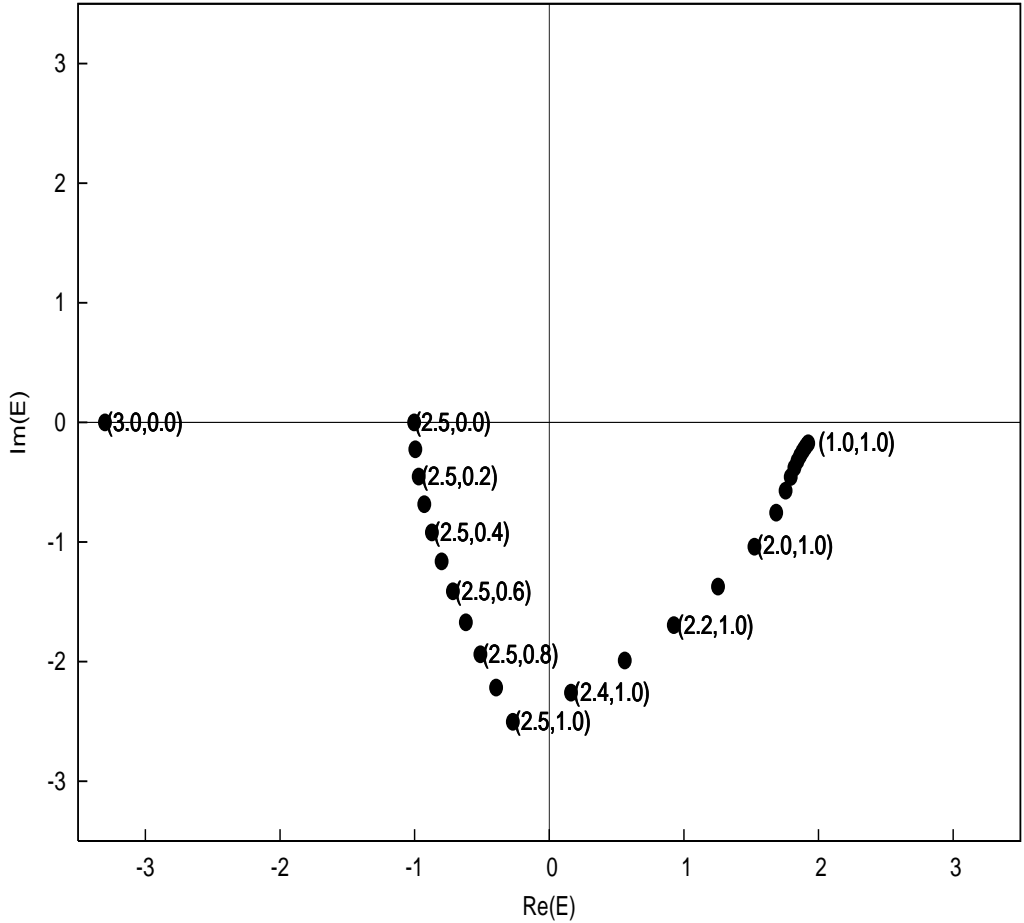}}} \par}
\vspace{0.3cm}
\caption{
Energy trajectory in the complex energy plane for the \( \Sigma -\alpha  \)
potential. The numbers at each point indicate the multiplicative factors
by which the attractive real part and the over all imaginary parts
were multiplied.} \label{fig1}
  \end{center}
\end{figure}

Before we present the result for the three-body system \( \Sigma \Sigma \alpha  \)
we firstly investigate the properties of the underlying two-body sub-systems.
The complex \( \Sigma -\alpha  \) potential leads to a complex energy
eigenvalue. We locate the one with the {}``lowest'' energy in the
following manner. We neglect the imaginary part and multiply the attractive
real part by some enhancement factor. Choosing for instance that factor
to be 2.5 we find a binding energy of -1.0 MeV. Next we allow the
imaginary part to increase from 0 in steps of 0.1 until we reach the
physical value 1. In this manner we find the energy trajectory in
the complex energy plane shown in Fig.1. We end up with the complex
energy position in the lower energy half plane just below the unitarity
cut from 0 to infinity. To reach that final position we have chosen
that detour in the complex energy plane which appears to us more feasible
than a direct energy search for the physical potential strength. Thus
we find that the effective \( \Sigma -\alpha  \) potential is not
strong enough to generate a complex energy with a negative real part.
The energy search in the complex energy plane was greatly simplified
by using a method of analytical continuation in the form of the point
method \cite{ref16}.

\begin{figure}
\begin{center}
\vspace{0.3cm}
{\centering \resizebox*{!}{0.9\textheight}{\includegraphics{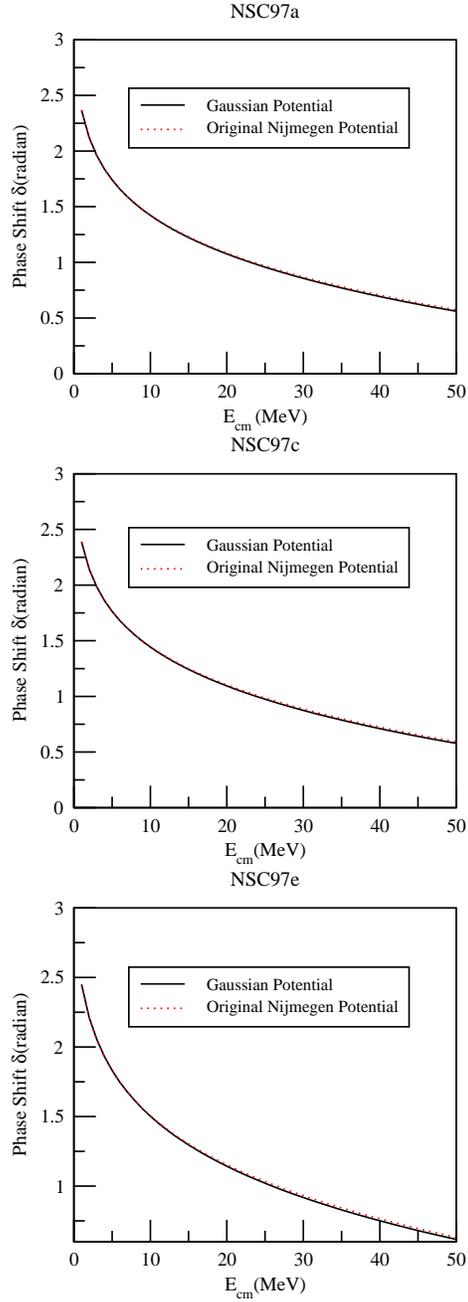}} \par}
\vspace{0.3cm}
\caption{
Comparison of \( \Sigma \Sigma  \)-phase shifts for the
original Nijmegen potentials NSC97a,c,e and the Gaussian potentials
for \( ^{1}S_{0} \) \( T=2 \).}
\label{fig2}
  \end{center}
\end{figure}

The chosen \( \Sigma -\Sigma  \) potentials support a bound state.
The corresponding binding energies are displayed in Table \ref{table3} for the
original Nijmegen potentials and the simulated ones. The phase shifts
in the state \( ^{1}S_{0} \) and T=2 for the original and simulated
potentials agree perfectly well with each other as shown in Fig.2.
Though it is presumably unrealistic that two \( \Sigma  \)'s are
bound it is the result of the meson based Nijmegen potentials, whose
parameters have been fixed to very many data in the nucleon-nucleon
and hyperon-nucleon sectors and where SU(3) symmetry arguments allow
for a prediction to the S=-2 sector. We shall comment below on an
ad hoc weakening of those potentials.

\begin{table}[tp]
\begin{center}
\begin{tabular}{|c|c|c|}
\hline 
&
Gaussian&
Nijmegen\\
\hline
\hline 
NSC97a&
-2.253&
-2.250\\
\hline 
NSC97c&
-2.460&
-2.437\\
\hline 
NSC97e&
-3.214&
-3.122\\
\hline
\end{tabular}
\caption{
Comparison of the two-body \( \Sigma \Sigma  \) binding
energies in MeV for the original Nijmegen potentials NSC97a,c,e and
 the simulated Gaussian potentials.}
  \label{table3}
  \end{center}
\end{table}

We are interested to see the outcome of those dynamical assumptions
for the \( \Sigma \Sigma \alpha  \) system. In this exploratory investigation
we restrict all orbital angular momenta to be zero. In the first step
we neglect the imaginary part of the \( \Sigma -\alpha  \) potential.
It turns out that in all cases using the original Nijmegen \( \Sigma -\Sigma  \)
potentials or the simulated ones, the three-body system is bound.
The results are given in Table \ref{table4}. Then we switch on the imaginary
part of the \( \Sigma -\alpha  \) potential in steps of 0.1. The
results are displayed in Fig.3 and Fig.4 for the two cases. The effect
of the imaginary part in the \( \Sigma -\alpha  \) potential shifts
the real part of the energy slightly to the right and introduces a
small negative imaginary part. The resulting final energy positions
are displayed in Table \ref{table5}.

\begin{figure}
\begin{center}
\vspace{0.3cm}
{\centering \resizebox*{1\textwidth}{0.6\textheight}{\rotatebox{-90}{\includegraphics{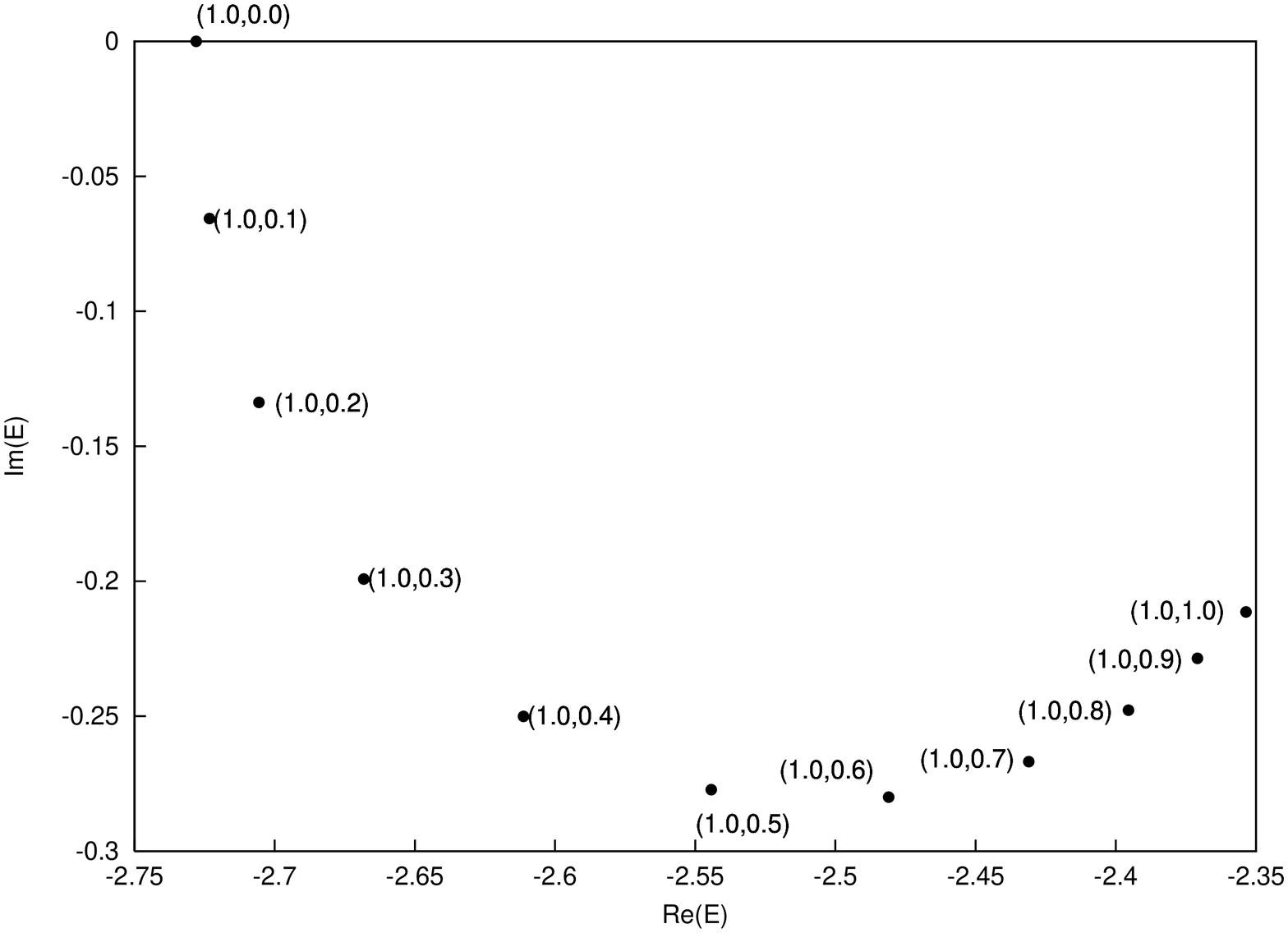}}} \par}
\vspace{0.3cm}
\caption{
Energy trajectories in the complex energy plane for the
\( ^{6}_{\Sigma \Sigma }He \) system using the original \( \Sigma -\Sigma  \)
Nijmegen NSC97e interaction. The numbers at each point indicate the
multiplicative factors by which the attractive real part and the over
all imaginary parts were multiplied.}
\label{fig3}
\end{center}
\end{figure}

\begin{figure}
\begin{center}
\vspace{0.3cm}
{\centering \resizebox*{1\textwidth}{0.6\textheight}{\rotatebox{-90}{\includegraphics{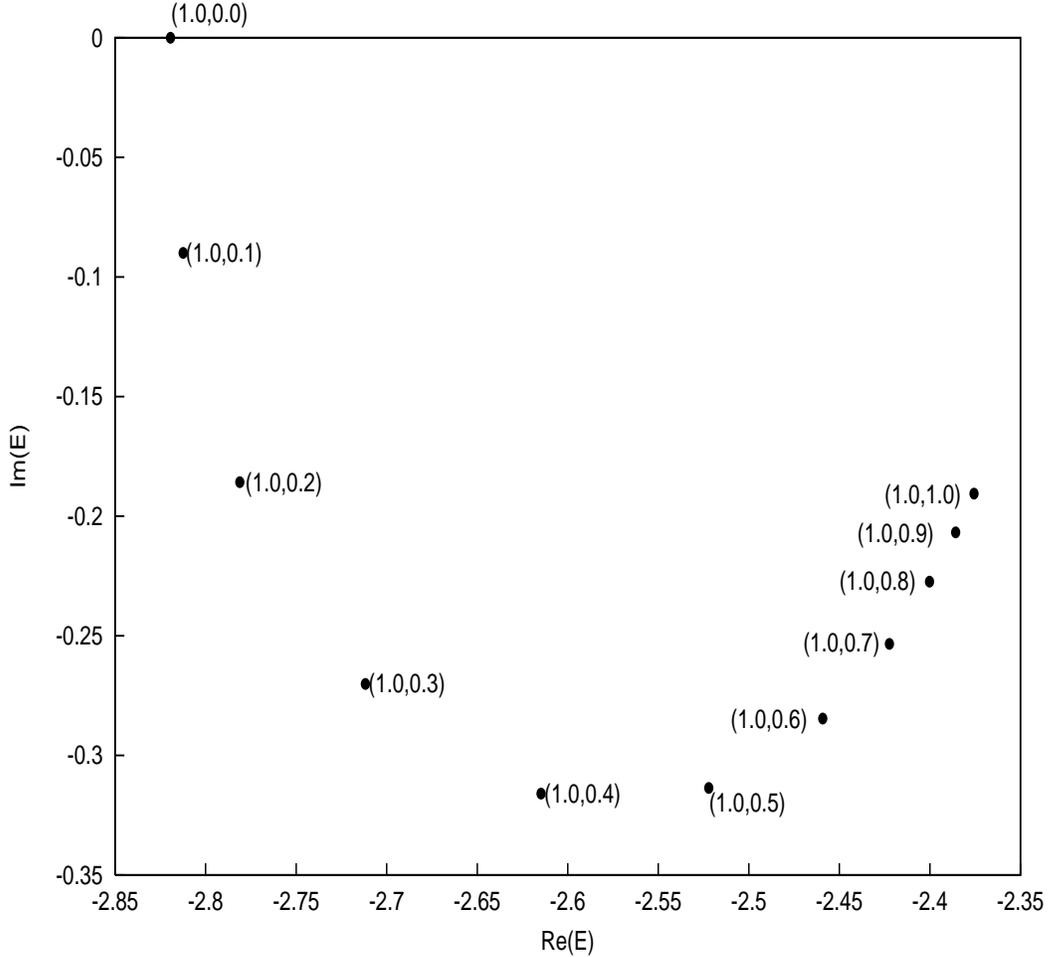}}} \par}
\vspace{0.3cm}
\caption{
Energy trajectories in the complex energy plane for the
\( ^{6}_{\Sigma \Sigma }He \) system using the simulated Gaussian
form for the \( \Sigma -\Sigma  \) Nijmegen NSC97e interaction. The
numbers at each point indicate the multiplicative factors by which
the attractive real part and the overall imaginary parts were multiplied. 
}\label{fig4}
\end{center}
\end{figure}

\begin{table}[tp]
\begin{center}
\begin{tabular}{|c|c|c|c|}
\hline 
&
NSC97a&
NSC97c&
NSC97e\\
\hline
\hline 
Nijmegen&
-1.840&
-2.378&
-2.728\\
\hline 
Gaussian &
-1.841&
-2.400&
-2.819\\
\hline
\end{tabular}
\caption{
The energy eigen values in MeV for the \( \Sigma \Sigma \alpha  \)
system using the simulated Gaussian and the original Nijmegen \( \Sigma -\Sigma  \)
potentials together with the real part alone of the \( \Sigma -\alpha  \)
potential.}
  \label{table4}
  \end{center}
\end{table}

\begin{table}[tp]
\begin{center}
\begin{tabular}{|c|c|c|c|}
\hline 
&
NSC97a&
NSC97c&
NSC97e\\
\hline
\hline 
Nijmegen&
-1.418-~i0.202&
-2.34-~i0.014&
-2.376-~i0.191\\
\hline 
Gaussian&
-1.492-~i0.218&
-2.323-~i0.017&
-2.354-~i0.211\\
\hline
\end{tabular}
\caption{
 The complex energy eigen values in MeV for the \( \Sigma \Sigma \alpha  \)
system using the simulated Gaussian and the original Nijmegen \( \Sigma -\Sigma  \)
potentials together with the complex \( \Sigma -\alpha  \) potentials.}
  \label{table5}
  \end{center}
\end{table}

In order to explore the outcome using \( \Sigma -\Sigma  \) potentials
which are weaker than the ones we used, we multiplied them by overall
factors 0.9 and 0.8. The resulting \( \Sigma -\Sigma  \) binding
energies for the original NSC97e potential are -1.5 and -0.4 MeV,
respectively ( in comparison to the original value of -3.1 MeV). Then
we performed again the energy search for the \( \Sigma -\Sigma -\alpha  \)
system starting with zero imaginary part for the effective \( \Sigma -\alpha  \)
potential. In case of the reduction factor 0.9 (0.8) this leads to
a 3-body binding energy of -1.08 MeV (-0.32 MeV). For the full imaginary
part we end up with (0.68 - i 0.21) MeV for the factor 0.9 and with(-0.279
- i 2.0\( \times  \) \( 10^{-2} \) ) MeV for the factor 0.8. We
conjecture that a low lying quasi bound state or a resonance for the
\( \Sigma -\Sigma -\alpha  \) system in the T=2 might exist in reality. 

Finally we would like to mention a possible means to approach such
a state experimentally. One could think of the two step process of
\( (K^{-},K^{+}) \) reaction on a \( ^{6}Li \) target;

\( K^{-}+^{6}Li\rightarrow \Sigma ^{0}+\pi ^{0}+n+^{4}_{2}He \)

\( \pi ^{0}+n\rightarrow \Sigma ^{-}+K^{+} \)

In this manner one populates two \( \Sigma  \)'s together with \( \alpha  \).
It remains of course the task to estimate the reaction rates.

\section{Summary and Outlook}
\label{sec4}
We developed a Faddeev code for the effective three-body systems \( \Lambda \Lambda \alpha  \)
and \( \Sigma \Sigma \alpha  \). It is formulated in momentum space
and is applicable for any type of two-baryon forces. This allowed
us to use directly the original Nijmegen forces, which is the first
time, to the best of our knowledge, that they have been applied for
these systems. We tested our code in a bench mark model study for
the \( \Lambda \Lambda \alpha  \) system reproducing perfectly well
the results from two earlier studies. Our results for the \( \Sigma \Sigma \alpha  \)
system lead to a quasi bound state with a small negative imaginary
part. The negative real part of the energy ranges between -1.4 and
-2.4 MeV. These numbers are based on the original or simulated Nijmegen
potentials for the \( \Sigma \Sigma  \) system in the state T=2,
which support a bound state with a binding energy of about -2.5 MeV.
Further we use a \( \Sigma -\alpha  \) optical model potential, which
by itself supports a complex energy eigen value of about (-2~-i 0.1)
MeV. We also artificially reduced the overall strength of the \( \Sigma -\Sigma  \)
potential by factors 0.9 and 0.8, which moved the 3-body \( \Sigma -\Sigma -\alpha  \)
energies towards -0.68 and -0.279 MeV, respectively,with small widths.

Both dynamical assumptions on the \( \Sigma -\Sigma  \) and \( \Sigma -\alpha  \)
potentials should be critically reinvestigated in the future. Upcoming
meson based \( \Sigma -\Sigma  \) potentials without a bound state
should be used and in addition the effective \( \Sigma -\alpha  \)
potential should be generated more consistently using realistic \( \alpha  \)
particle wave functions in conjunction with \( \Sigma  \)-nucleon
forces related to the same theoretical model as for the \( \Sigma -\Sigma  \)
interaction. A low lying state for the \( \Sigma -\Sigma -\alpha  \)
system with isospin T=2 would provide interesting additional information
on the dynamics in the strangeness S=-2 sector.

\section*{Acknowledgments}

Htun Htun Oo would like 
to express his gratitude towards DAAD ( Deutscher Akademischer
Austauschdienst) for financial
support in the framework of the  Sandwich Scheme. K.S. Myint acknowledges
her thanks to Professor Y.Akaishi for his fruitful discussions. W.Gl\"ockle
thanks the Department of Physics, Mandalay University, for the very
kind hospitality extended to him during his stay there when this work
has been completed.

\section*{Appendix}

The purely geometrical coefficients occuring in Eqs(\ref{eq.19}) and (\ref{eq.20})  are 

\begin{eqnarray}
&&\mathcal{G}_{\alpha _{3}\alpha '_{1}}\left( q_{3},q'_{1},x\right)   
\cr 
&=&  \frac{1}{4}(-)^{j+\lambda +1+l'}\sqrt{\hat{j}\, \hat{j}'\, \hat{I}'\, \hat{s}\, \hat{l}\, \hat{\lambda }\, \hat{l}'\, \hat{\lambda }'}\, \sum _{L s}\hat{L}(-)^{L}\left\{ \begin{array}{ccc}
s & l & j\\
\lambda  & J & L
\end{array}\right\} \left\{ \begin{array}{ccc}
l' & \frac{1}{2} & j'\\
\lambda ' & \frac{1}{2} & I'\\
L & s & J
\end{array}\right\} \nonumber \\
 &  & \sum _{l_{1}+l_{2}=l}\, \sum _{l'_{1}+l'_{2}=l'}\rho ^{l_{2}}_{1}\, \rho ^{l'_{1}}_{2}\, \left( q_{3}\right) ^{l_{2}+l'_{2}}\, \left( q'_{1}\right) ^{l'_{1}+l_{1}}\nonumber \\
 &  & \sqrt{\frac{\left( 2l+1\right) !}{2l_{1}!2l_{2}!}}\, \sqrt{\frac{\left( 2l'+1\right) !}{2l'_{1}!2l'_{2}!}}\, \sum _{f}\left\{ \begin{array}{ccc}
l_{1} & l_{2} & l\\
\lambda  & L & f
\end{array}\right\} C\left( l_{2}\lambda f;00\right) \nonumber \\
 &  & \sum _{f'}\left\{ \begin{array}{ccc}
l'_{2} & l'_{1} & l'\\
\lambda ' & L' & f'
\end{array}\right\} C\left( l'_{1}\lambda 'f';00\right) \sum _{k}\, \hat{k}\, P_{k}(x)\, \sum _{h}\, \hat{h}\, g_{h}\sum _{f''}\, C\left( khf'';00\right) ^{2}\nonumber \\
 &  & \left\{ \begin{array}{ccc}
f & l_{1} & L\\
f' & l'_{2} & f''
\end{array}\right\} C\left( f''l'_{2}f;00\right) C\left( f''l_{1}f';00\right) \nonumber \\
 &  & \nonumber 
\end{eqnarray}

with

\[
g_{h}=\int ^{1}_{-1}dxP_{h}(x)\frac{1}{\left| \rho _{2}\mathbf{q}'_{1}+\mathbf{q}_{3}\right| ^{l'}}\frac{1}{\left| \mathbf{q}'_{1}+\rho _{1}\mathbf{q}_{3}\right| ^{l}}\]

and 

\begin{eqnarray}
&&\mathcal{G}_{\alpha _{1}\alpha '_{1}}\left( q_{1},q'_{1},x\right) 
\cr
 & = & \frac{1}{4}\sum _{LS}\, (-)^{1-S}\sqrt{\hat{j}\, \hat{I}\, \hat{j}'\, \hat{I}'\hat{l}\, \hat{\lambda }\hat{l}'\, \hat{\lambda }'}\, \, \hat{L}\, \hat{S}\left\{ \begin{array}{ccc}
l & \frac{1}{2} & j\\
\lambda  & \frac{1}{2} & I\\
L & S & J
\end{array}\right\} \left\{ \begin{array}{ccc}
l' & \frac{1}{2} & j'\\
\lambda ' & \frac{1}{2} & I'\\
L & S & J
\end{array}\right\} \nonumber \\
 &  & \sum _{l_{1}+l_{2}=l}\, \sum _{l'_{1}+l'_{2}=l'}\, \rho ^{l'_{1}+l_{2}}\, \, \left( q'_{1}\right) ^{l_{1}+l'_{1}}\, \left( q_{1}\right) ^{l_{2}+l'_{2}}\nonumber \\
 &  & \sqrt{\frac{\left( 2l+1\right) !}{2l_{1}!2l_{2}!}}\, \sqrt{\frac{\left( 2l'+1\right) !}{2l'_{1}!2l'_{2}!}}\, \sum _{f}\left\{ \begin{array}{ccc}
l_{1} & l_{2} & l\\
\lambda  & L & f
\end{array}\right\} C\left( l_{2}\lambda f;00\right) \nonumber \\
 &  & \sum _{f'}\left\{ \begin{array}{ccc}
l'_{2} & l'_{1} & l'\\
\lambda ' & L & f'
\end{array}\right\} C\left( l'_{1}\lambda 'f';00\right) \sum _{k}\hat{k}P_{k}(x)\, \sum _{h}\, \hat{h}g_{h}\nonumber \\
 &  & \sum _{f''}\, C\left( khf'';00\right) ^{2}\left\{ \begin{array}{ccc}
f & l_{1} & L\\
f' & l'_{2} & f''
\end{array}\right\} C\left( f''l'_{2}f;00\right) C\left( f''l_{1}f';00\right) \nonumber \\
 &  & \nonumber 
\end{eqnarray}

with \[
g_{h}=\int ^{1}_{-1}dxP_{h}(x)\frac{1}{\left| \rho \mathbf{q}_{1}+\mathbf{q}'_{1}\right| ^{l}}\frac{1}{\left| \mathbf{q}_{1}+\rho \mathbf{q}'_{1}\right| ^{l'}}\]


%

\end{document}